\documentclass[12pt,a4paper,fleqn]{article}
\begin{document}

\title{Integrability of a Generalized Ito System:\\the Painlev\'{e} Test}
\author{Ay\c{s}e \textsc{Karasu (Kalkanli)}\thanks{E-mail:
\textit{akarasu@metu.edu.tr}} , Atalay \textsc{Karasu}\thanks{E-mail:
\textit{karasu@metu.edu.tr}}
\and and Sergei Yu.\ \textsc{Sakovich}\thanks{Permanent address: \textit{Institute
of Physics, National Academy of Sciences, 220072 Minsk, Belarus}. E-mail:
\textit{saks@pisem.net}}\bigskip\\\textit{Department of Physics,
Middle East Technical University,}\\\textit{06531 Ankara, Turkey}}
\date{}
\maketitle

\begin{abstract}
It is shown that a generalized Ito system of four coupled nonlinear evolution
equations passes the Painlev\'{e} test for integrability in five distinct
cases, of which two were introduced recently by Tam, Hu and Wang. A conjecture
is formulated on integrability of a vector generalization of the Ito
system.\medskip\newline \textsf{KEYWORDS: integrable systems, Painlev\'{e}
analysis}\bigskip

\end{abstract}

Recently, Tam, Hu and Wang \cite{THW} introduced the following two systems of
coupled nonlinear evolution equations:%
\begin{equation}%
\begin{array}
[c]{l}%
u_{t}=v_{x},\\
v_{t}=-2v_{xxx}-6\left(  uv\right)  _{x}-12ww_{x}+6p_{x},\\
w_{t}=w_{xxx}+3uw_{x},\\
p_{t}=p_{xxx}+3up_{x},
\end{array}
\label{ito1}%
\end{equation}
and%
\begin{equation}%
\begin{array}
[c]{l}%
u_{t}=v_{x},\\
v_{t}=-2v_{xxx}-6\left(  uv\right)  _{x}-6\left(  wp\right)  _{x},\\
w_{t}=w_{xxx}+3uw_{x},\\
p_{t}=p_{xxx}+3up_{x},
\end{array}
\label{ito2}%
\end{equation}
which are generalizations of the well-known integrable Ito system \cite{Ito}%
\begin{equation}%
\begin{array}
[c]{l}%
u_{t}=v_{x},\\
v_{t}=-2v_{xxx}-6\left(  uv\right)  _{x}.
\end{array}
\label{ito}%
\end{equation}
Hirota bilinear representations of the systems (\ref{ito1}) and (\ref{ito2}),
4-soliton solutions of the system (\ref{ito1}) with $p=0$, and 3-soliton
solutions of the system (\ref{ito2}) were found in ref.~\cite{THW}. More
recently, a new type of 3-soliton solutions with constant boundary conditions
at infinity was found for the system (\ref{ito1}) with $p=0$ in
ref.~\cite{TMHW}. The question on integrability of the systems (\ref{ito1})
and (\ref{ito2}), in the sense of existence of Lax pairs, infinitely many
conservation laws, and N-soliton solutions, was posed in ref.~\cite{THW}.

In this Letter, we show that the Tam-Hu-Wang systems (\ref{ito1}) and
(\ref{ito2}) must be integrable according to positive results of the
Painlev\'{e} test. We apply the Painlev\'{e} test for integrability to the
following generalized Ito system:%
\begin{equation}%
\begin{array}
[c]{l}%
u_{t}=v_{x},\\
v_{t}=-2v_{xxx}-6\left(  uv\right)  _{x}+aww_{x}+bpw_{x}+cwp_{x}%
+dpp_{x}+fw_{x}+gp_{x},\\
w_{t}=w_{xxx}+3uw_{x},\\
p_{t}=p_{xxx}+3up_{x},
\end{array}
\label{gito}%
\end{equation}
where $a,b,c,d,f,g$ are arbitrary constants. The system (\ref{gito}) turns out
to possess the Painlev\'{e} property under the only constraint imposed on its
coefficients: $c=b$. Then, using affine transformations of $w$ and $p$, we
reduce the system (\ref{gito}) with $c=b$ to five distinct cases, of which two
being eqs.~(\ref{ito1}) and (\ref{ito2}). Finally, we propose a conjecture on
integrability of a multi-component generalization of the Tam-Hu-Wang systems
(\ref{ito1}) and (\ref{ito2}).

Let us apply the Painlev\'{e} test for integrability to the system
(\ref{gito}), following the so-called Weiss-Kruskal algorithm of singularity
analysis \cite{WTC}, \cite{JKM}. The system (\ref{gito}) is a normal system of
four partial differential equations of total order ten, and its general
solution must contain ten arbitrary functions of one variable. A hypersurface
$\phi\left(  x,t\right)  =0$ is non-characteristic for this system if
$\phi_{x}\phi_{t}\neq0$, and we set $\phi_{x}=1$ without loss of generality.
The substitution of the expansions%
\begin{equation}%
\begin{array}
[c]{l}%
u=u_{0}\left(  t\right)  \phi^{\alpha}+\cdots+u_{n}\left(  t\right)
\phi^{n+\alpha}+\cdots,\\
v=v_{0}\left(  t\right)  \phi^{\beta}+\cdots+v_{n}\left(  t\right)
\phi^{n+\beta}+\cdots,\\
w=w_{0}\left(  t\right)  \phi^{\gamma}+\cdots+w_{n}\left(  t\right)
\phi^{n+\gamma}+\cdots,\\
p=p_{0}\left(  t\right)  \phi^{\delta}+\cdots+p_{n}\left(  t\right)
\phi^{n+\delta}+\cdots
\end{array}
\label{exp}%
\end{equation}
into the system (\ref{gito}) determines the branches, i.e.\ the admissible
dominant behavior of solutions (values of $\alpha,\beta,\gamma,\delta
,u_{0},v_{0},w_{0},p_{0}$) and the corresponding positions $n$ of the
resonances (where arbitrary functions can appear in the expansions (\ref{exp})).

The system (\ref{gito}) admits many branches, but the presence or absence of
most of them depends on values of the parameters $a,b,c,d,f,g$. For this
reason, it is useful to start the analysis from the following singular branch:%
\begin{equation}%
\begin{array}
[c]{l}%
\alpha=\beta=-2,\quad\gamma=\delta=-1,\\
u_{0}=-2,\quad v_{0}=-2\phi_{t},\quad\forall\,w_{0}\left(  t\right)
,\quad\forall\,p_{0}\left(  t\right)  ,\\
n=-1,0,0,1,1,2,4,5,5,6,
\end{array}
\label{gb}%
\end{equation}
which is admitted by the system (\ref{gito}) at any choice of its parameters.
According to the positions of resonances, the branch (\ref{gb}) is generic,
i.e. it represents the general solution of the system (\ref{gito}). Now,
constructing the recursion relations for the coefficients of the expansions
(\ref{exp}), and checking the consistency of those recursion relations at the
resonances of the branch (\ref{gb}), we obtain the compatibility condition
$\left(  b-c\right)  \left(  w_{0,t}p_{0}-w_{0}p_{0,t}\right)  =0$ at $n=5$.
If $c\neq b$, some logarithmic terms must be introduced into the expansions
(\ref{exp}). Therefore the system (\ref{gito}) can possess the Painlev\'{e}
property only if $c=b$. Setting $c=b$ hereafter, we find that the recursion
relations are consistent at all the resonances of the branch (\ref{gb}).

Before proceeding to other branches, let us notice that, in the case of $c=b$,
we can fix all the free parameters of the system (\ref{gito}) by means of the
affine transformation%
\begin{equation}%
\begin{array}
[c]{l}%
w\rightarrow\xi_{1}w+\xi_{2}p+\xi_{3},\\
p\rightarrow\xi_{4}w+\xi_{5}p+\xi_{6}%
\end{array}
\label{tr}%
\end{equation}
with appropriately chosen constants $\xi_{1},\ldots,\xi_{6}$, $\xi_{1}\xi
_{5}\neq\xi_{2}\xi_{4}$. Certainly, this transformation has no effect on the
presence or absence of the Painlev\'{e} property. If $ad\neq b^{2}$, then,
using the transformation (\ref{tr}), we can make%
\begin{equation}
b=-6,\quad a=d=f=g=0. \label{c1}%
\end{equation}
If $ad=b^{2},a\neq0,ag\neq bf$, or if $a=b=0,d\neq0,f\neq0$, we can make%
\begin{equation}
a=-12,\quad g=6,\quad b=d=f=0. \label{c2}%
\end{equation}
If $ad=b^{2},a\neq0,ag=bf$, or if $a=b=0,d\neq0,f=0$, we can make%
\begin{equation}
a=-12,\quad b=d=f=g=0. \label{c3}%
\end{equation}
If $a=b=d=0,f\neq0$, or if $a=b=d=f=0,g\neq0$, we can make%
\begin{equation}
a=b=d=f=0,\quad g=6. \label{c4}%
\end{equation}
And the case of%
\begin{equation}
a=b=d=f=g=0 \label{c5}%
\end{equation}
needs no transformation. These five cases (\ref{c1})--(\ref{c5}) of the system
(\ref{gito}) are not related to each other by the transformation (\ref{tr}).

Having reduced the system (\ref{gito}) with $c=b$ to the five distinct cases
(\ref{c1})--(\ref{c5}), we find that the following singular non-generic
branches must be studied as well:%
\begin{equation}%
\begin{array}
[c]{l}%
\alpha=\beta=\gamma=\delta=-2,\\
u_{0}=-4,\quad v_{0}=-4\phi_{t},\quad w_{0}p_{0}=-8\phi_{t},\quad
\forall\,w_{0}\left(  t\right)  \;\mathrm{or}\;\forall\,p_{0}\left(  t\right)
,\\
n=-2,-1,0,2,2,3,4,6,7,8
\end{array}
\label{b1}%
\end{equation}
and%
\begin{equation}%
\begin{array}
[c]{l}%
\alpha=\beta=-2,\quad\gamma=0,\quad\delta=-4,\\
u_{0}=-10,\quad v_{0}=-10\phi_{t},\quad w_{0}p_{0}=-80\phi_{t},\quad
\forall\,w_{0}\left(  t\right)  \;\mathrm{or}\;\forall\,p_{0}\left(  t\right)
,\\
n=-5,-4,-1,0,2,4,6,7,8,12
\end{array}
\label{b2}%
\end{equation}
in the case (\ref{c1});%
\begin{equation}%
\begin{array}
[c]{l}%
\alpha=\beta=-2,\quad\gamma=0,\quad\delta=-4,\\
u_{0}=-10,\quad v_{0}=-10\phi_{t},\quad p_{0}=80\phi_{t},\quad\forall
\,w_{0}\left(  t\right)  ,\\
n=-5,-4,-1,0,2,4,6,7,8,12
\end{array}
\label{b3}%
\end{equation}
in the case (\ref{c2});%
\begin{equation}%
\begin{array}
[c]{l}%
\alpha=\beta=\gamma=\delta=-2,\\
u_{0}=-4,\quad v_{0}=-4\phi_{t},\quad w_{0}^{2}=-8\phi_{t},\quad\forall
\,p_{0}\left(  t\right)  ,\\
n=-2,-1,0,2,2,3,4,6,7,8
\end{array}
\label{b4}%
\end{equation}
in the cases (\ref{c2}) and (\ref{c3}); and%
\begin{equation}%
\begin{array}
[c]{l}%
\alpha=\beta=-2,\quad\gamma=\delta=-4,\\
u_{0}=-10,\quad v_{0}=-10\phi_{t},\quad p_{0}=80\phi_{t},\quad\forall
\,w_{0}\left(  t\right)  ,\\
n=-5,-1,0,2,4,4,6,8,11,12
\end{array}
\label{b5}%
\end{equation}
in the case (\ref{c4}). Then, using the \textit{Mathematica} system
\cite{Wol}, we prove that the recursion relations are consistent at the
resonances of the branches (\ref{b1})--(\ref{b5}), and therefore no
logarithmic terms should be introduced into the expansions (\ref{exp}).

Now, we can conclude that the generalized Ito system (\ref{gito}) passes the
Painlev\'{e} test for integrability if, and only if, $c=b$, or, up to the
equivalence (\ref{tr}), in the five distinct cases (\ref{c1})--(\ref{c5}). The
cases (\ref{c1}) and (\ref{c2}) correspond to the Tam-Hu-Wang systems
(\ref{ito2}) and (\ref{ito1}), respectively.

The obtained results of the singularity analysis are highly suggestive that
the system (\ref{gito}) with $c=b$ must be integrable in the Lax sense.
Moreover, we \textit{conjecture} that, for any constant $k_{ij}$ and $l_{i}$,
$i,j=1,\ldots,m$, and any integer $m$, the following system of $m+2$ coupled
nonlinear evolution equations for $u,v,q_{1},\ldots,q_{m}$, a vector
generalization of the Ito system (\ref{ito}),%
\begin{equation}%
\begin{array}
[c]{l}%
u_{t}=v_{x},\\
v_{t}=-2v_{xxx}-6\left(  uv\right)  _{x}+\left(  \sum_{i,j}k_{ij}q_{i}%
q_{j}+\sum_{i}l_{i}q_{i}\right)  _{x},\\
q_{i,t}=q_{i,xxx}+3uq_{i,x},\quad i=1,\ldots,m,
\end{array}
\label{vito}%
\end{equation}
passes the Painlev\'{e} test for integrability and possesses a parametric
zero-curvature representation.\bigskip

The authors are grateful to the Scientific and Technical Research Council of
Turkey (T\"{U}B\.{I}TAK) for support.

\end{document}